\documentclass[11pt,reqno,final]{article}

\usepackage{amsmath,amsfonts,amssymb,amsthm,version}
\usepackage{mathrsfs,fancybox,pifont}
\usepackage{graphicx}
\usepackage{algorithm,algorithmic}
\usepackage{cite}
\usepackage{setspace}
\usepackage{url,hyperref}
\usepackage{fullpage} % set 1 inch margins
\usepackage[notcite,notref]{showkeys}
\usepackage{color}
\usepackage{subfigure,multirow}
\usepackage{epstopdf}
\usepackage{cases}
\usepackage{geometry}
\usepackage{mathtools}
\usepackage{authblk}
\usepackage{lipsum}
\usepackage{indentfirst}

\allowdisplaybreaks

\numberwithin{equation}{section}
\theoremstyle{plain}
\newtheorem{theorem}{Theorem}%[section]
\newtheorem{proposition}{Proposition}%[section]
\newtheorem{lemma}{Lemma}%[section]
\newtheorem{corollary}{Corollary}%[section]

\theoremstyle{definition}

\theoremstyle{remark}
\newtheorem{remark}{Remark}%[section]

\title{Sharp Estimates for Optimal Multistage Group Partition Testing}
\author[1,2]{Guojiang Shao \\ \texttt{gjshao@yandex.com}}
\affil[1]{School of Mathematical Sciences, Fudan University, Shanghai 200433, China}
\affil[2]{School of Mathematical Science, Zhejiang University, Hangzhou 310058, China}
\date{}

\begin{document}
\maketitle

\begin{abstract}

  % In multistage group testing, the tests within the same stage are considered nonadaptive, while those conducted across different stages are adaptive. Specifically, when the pools within the same stage are disjoint, meaning that the entire set is divided into several disjoint subgroups, it is referred to as a multistage group partition testing problem, denoted as the $(n, d, s)$ problem, where $n$, $d$, and $s$ represent the total number of items, defectives, and stages respectively. The optimal number of tests required for this problem is denoted as $T_d(n, s)$. This paper presents exact solutions for the $(n,1,s)$ and $(n,d,2)$ problems for the first time. Additionally, a general dynamic programming approach is developed for the $(n,d,s)$ problem. Significantly it is proved that $\lceil ds\bigl( \frac{n}{d} \bigr)^\frac{1}{s} \rceil \leq T_d(n,s) \leq \lceil ds\bigl( \frac{n}{d} \bigr)^\frac{1}{s} \rceil + 1$ for large $n$, where both the upper and lower bounds are sharp for $T_d(n,s)$. If the defective number in unknown but bounded, we can provide an algorithm with an optimal competitive ratio in the asymptotic sense. While assuming the prior distribution of the defective items, we also establish a well performing upper and lower bound estimates to the expectation of optimal strategy.

  In multistage group testing, the tests within the same stage are considered nonadaptive, while those conducted across different stages are adaptive. Especially, when the pools within the same stage are disjoint, meaning that the entire set is divided into several disjoint subgroups, it is referred to as a multistage group partition testing problem, denoted as the $(n, d, s)$ problem, where $n$, $d$, and $s$ represent the total number of items, defectives, and stages respectively. This paper presents exact solutions for the $(n,1,s)$ and $(n,d,2)$ problems for the first time. Furthermore, we develop a general dynamic programming framework for the $(n,d,s)$ problem, which allows us to derive the sharp estimation of upper and lower bounds.
  %If the defective number in unknown but bounded, we can provide an algorithm with an optimal competitive ratio in the asymptotic sense. While assuming the prior distribution of the defective items, we also establish a well performing upper and lower bound estimates to the expectation of optimal strategy.

\end{abstract}

% \newpage

%%%%%%%%%%%%%%%%%%%%%%%%%%%%%%%%%%%%%%%%%%%%%%%%%%%%%%%%%%%%%%%%%%%%%%%%%%%%%%%%%%
\section{Introduction and Results Review}

  The origin of group testing can be traced back to Dorfman \cite{dorfman1943detection}. The paper by Wolf \cite{wolf1985born} provided a historical overview of early developments, emphasizing adaptive algorithms. Du and Hwang \cite{du2000combinatorial,hwang2006pooling} gave extensive background on group testing, especially on adaptive testing, zero-error nonadaptive testing and related applications. Aldridge et al. \cite{aldridge2019group} surveyed recent developments in the group testing problem from an information-theoretic perspective. More broadly, group testing has garnered significant attention in recent years, as demonstrated by works such as \cite{d2014lectures,malyutov2013search,kautz1964nonrandom,katona1973combinatorial,chan2011non}.

  Group testing questions can generally be divided into two categories: the combinatorial group testing (CGT) and the probabilistic group testing (PGT). In probabilistic models, the defective items are assumed to follow some probability distribution and its aim is to minimise the expectation of tests needed to identify all defective items. While in combinatorial group testing, the goal is to minimise the number of tests needed in a worst-case scenario, that is, creating a minmax algorithm and no knowledge of the distribution of defectives is assumed. In this article, we focus exclusively on group testing problems within the combinatorial framework.

  In classical combinatorial group testing model, if we consider $n$ items containing exactly $d$ defectives, and error-free detection is employed to identify these defective individuals, there are only two possible outcomes for any subset detection: positive and negative. A positive result indicates that there is at least one defective individual in the subset, while a negative result indicates that there is no defective individual in the subset. This problem is sometimes referred to as the hypergeometric group testing problem and is commonly denoted by $(d,n)$ problem (see e.g. Du et al.\cite{du2000combinatorial}).

  There are two general types of group testing algorithms: adaptive (or sequential) algorithms and nonadaptive algorithms. In adaptive algorithms, tests are conducted one by one, and the results of previous tests are assumed to be known at the time of current test. In nonadaptive algorithms, no previous testing information is known, and all tests are specified simultaneously. A compromise between the two types of algorithms is realized by the so-called multistage algorithm, for a multistage algorithm, tests within the same stage are considered nonadaptive, while those across different stages are considered adaptive. That is, the test can use the results of previous stages, but results from the same stage cannot be used. If the number of stages in a multistage algorithm is fixed to $s$, it can be called an $s$-stage algorithm. Specifically in the context of multistage group testing, an element may participate in several pools within a stage, but if we always divide the whole into a disjoint union of several subgroups, the multistage algorithm can be simplified, we call this model as multistage group partition testing (MSGPT) which is the core model of our discussions in the following text (see e.g. Gajpal et al.\cite{gajpal2022optimal}). Obviously, the MSGPT problem is a special type of multistage group testing problem, and the multistage algorithm is a special case of adaptive algorithm.

  % When considering more realistic settings or if some assumptions are not perfect, we can derive distinct types of group testing problems. Aldridge et al.\cite{aldridge2019group} enumerated various assumptions on the mathematical model used, including adaptive vs. nonadaptive, zero error probability vs. small error probability, Binary vs. nonbinary outcomes, noiseless vs. noisy testing, known vs. unknown number of defectives and combinatorial vs. i.i.d. prior. For mathematical convenience, we usually consider the scenario where the number of defectives is fixed or at least bounded, and the defective set is uniformly random among all sets of this size. Alternatively one could explore the situations that the number of defectives is unknown or each item is defective independently with the same fixed low probability $p$ (which is called the i.i.d. prior), see for example, Aldridge et al.\cite{aldridge2019group}. More generally, we assume that each probability is small
  % and items are independent, but not necessarily identically distributed or we can consider giving a distribution of the defective numbers (e.g. Li et al. \cite{li2014group}). When the testing process is not exact and is thus subject to noise, Scarlett \cite{scarlett2018noisy} studied noise adaptive group testing and Scarlett et al. \cite{scarlett2020noisy} later examined the noise nonadaptive group testing. Additionally, Aldridge et al. \cite{aldridge2019group} provided a comprehensive review of group testing in the presence of noise.

  When we use a certain algorithm $A$ to detect the $(d, n)$ problem, we record the number of detections required by $M_A(d, n)$, and define
  \begin{equation*}
    M(d, n):=\min_A M_A(d, n).
  \end{equation*}
  One of the goals of the CGT problem is to solve $M(d,n)$. It is easy to see that for an adaptive algorithm, $M(1,n) = \lceil \log_2 n \rceil$, but for a general adaptive algorithm, $M(d,n)$ (where $d \geq 2$) is still an open question (see e.g. Du et al.\cite{du2000combinatorial}). Under an adaptive algorithm, for the $(d,n)$ problem, Hwang\cite{hwang1972method} proved that it takes at most $d-1$ tests more than the information lower bound of $\lceil \log_2 \tbinom{n}{d} \rceil$(here $\tbinom{n}{d} := \frac{n!}{d!(n-d)!}$) to find out all defectives using the generalized binary search. And Allemann \cite{allemann2013efficient} improved upon this result by proving that when $\frac{n}{d}\geq 38$ or $d\geq 10$, it is possible to find out all defectives with at most $0.187d+0.5\log_2(d)+5.5$ tests more than the information lower bound $\lceil \log_2 \tbinom{n}{d} \rceil$. As such, this kind of problem has essentially been solved, with little  room for further improvement.

  For a multistage algorithm, Li \cite{li1962sequential} extended Dorfman's two-stage algorithm (for PGT) to $s$ stages and demonstrated that $O\left( ds( \frac{n}{d} )^{1/s}\right)$ tests suffice.
  % that finding no more than $d$ defectives among $n$ items required no more than $\frac{e}{\log (e)} d \log (\frac{n}{d})$ tests. 
  Dyachkov et al. \cite{d1982bounds} proved that for a non-adaptive algorithm (i.e., a multi-stage algorithm with only one stage), at least
  \begin{equation*}
    \frac{d^2}{2 \log _2(e(d+1) / 2)} \log _2 n\bigl(1+o(1) \bigr),
  \end{equation*}
  tests are needed. When $s=2$, De Bonis et al. \cite{de2005optimal} proved that $O(d\log_2 n)$ tests are sufficient. In particular, in the strict multi-stage group testing setting with $d=2$ and $s=2$, Damaschke et al. \cite{damaschke2014strict} proved that
  \begin{equation*}
    t(n,2,2) \leq 2.44 \log_2 n+o(\log_2 n),
  \end{equation*}
  where $t(n,2,2)$ represented the optimal testing numbers in this case. And Damaschke et al. \cite{damaschke2013two} constructed the optimal random strategy for $d=1$ and $s \leq 2$. For $d=2$ and $s=4$, D'yachkov et al. \cite{d2016hypergraph} proved that $2 \log_2 n\bigl(1+o(1)\bigr)$ tests are sufficient. We can find only for certain special situations can there be some better results.

  %While if we concentrate on the MSGPT (thus simplified multistage group testing) problem, it's more realistic and more operable.
  In the background of MSGPT problem, for the case of $d=1$, Gajpal et al. \cite{gajpal2022optimal} borrowed the clustering idea to impose constraints on each stage, and proposed dynamic programming to characterize the optimal number of tests. However, they just gave some numerical results and we will give the solution by solving its corresponding dual problem. Li \cite{li1962sequential} proved that $O\left(d s( \frac{n}{d} )^{1/s}\right)$ tests are enough, and Damaschke et al. \cite{damaschke2013two} further proved that any testing strategy for $d$ defectives and $s$ stages needs at least $(1-o(1)) d^{\prime}s\left( \frac{n}{d^\prime} \right)^{1/s}$ tests in the worst case, where $d=o\left(n^{1/s}\right)$ and $d^{\prime}:=d-1$. This implies that the complexity of the $(n,d,s)$ problem is essentially $\Theta\left(d s( \frac{n}{d} )^{1/s}\right)$. And when the number of defectives $d$ was not known beforehand, Damaschke et al. \cite{damaschke2013two} also proved that one can identify $d$ defectives using $\left(d (s-1)( \frac{n}{d} )^{1/(s-1)}\right)$ tests. It's obvious there is a gap between the upper bound estimation of Li \cite{li1962sequential} and lower bound estimation of Damaschke et al. \cite{damaschke2013two}, the main work of us is to improve the estimation with dynamic programming approach.

  The remainder of the paper is organized as follows. Section $2$ presents some notations and preliminaries; and Section $3$ provides the exact solution for two special cases of MSGPT problem. Section $4$ studies the optimal upper and lower bounds of general MSGPT problem and gives the sharp estimates. Finally Section $5$ summarizes our findings and provides directions for future research. In order to better present the core results, some complex proofs are included in the appendix.

%%%%%%%%%%%%%%%%%%%%%%%%%%%%%%%%%%%%%%%%%%%%%%%%%%%%%%%%%%%%%%%%%%%%%%%%%%
\section{Preliminaries and Notations}

  In the context of multistage group testing, the goal is to identify all $d$ defectives among $n$ items in $s$ stages. Specially, if all tests at the same stage are disjoint simultaneous, which means that each item is only contained in one subgroup after dividing, we define this group testing problem as multistage group partition testing (MSGPT) problem (see e.g. Gajpal et al.\cite{gajpal2022optimal}). We will focus on the MSGPT model in the rest of this paper.

  % If we need to search out exact known $d$ defectives among $n$ items in $s$ stages, we refer this problem as the $(n,d,s)$ problem. The optimal number of tests is denoted as $T_d(n,s)$ with the optimal algorithm, even if the optimal algorithm is unknown. If we divide the whole into a disjoint union of $m$ subgroups in the first stage and next take the optimal algorithm, correspondingly set the optimal testing numbers as $T^m_d(n,s)$, it is known that

  When we aim to precisely locate \(d\) defective items among \(n\) items within \(s\) stages, this scenario is designated as the \((n, d, s)\) problem. We say solving the \((n, d, s)\) problem means finding the optimal detection strategy. The optimal number of tests, denoted by \(T_d(n, s)\), represents the least number of tests required by the optimal algorithm in the worst case, even if the specific optimal algorithm is not yet known. If we initially partition the entire set into \(m\) disjoint subgroups and subsequently apply the optimal algorithm, the corresponding optimal number of tests is designated as \(T^m_d(n, s)\). It is known that
  \begin{equation*}
    T_d(n,s) = \min_m T^m_d(n,s).
  \end{equation*}

  The following general assumptions are made throughout the text:
  \begin{enumerate}
    % \item We assume that all test results are accurate, meaning that there are no false positives or false negatives.
    \item We assume that all testing results are accurate, with no false positives or false negatives occurring.
    % \item Testing in different stages is considered adaptive, while testing in the same stage is considered non-adaptive.
    \item Tests conducted across different stages are classified as adaptive, whereas tests performed within the same stage are deemed non-adaptive.
    % \item To ensure that the test results are meaningful, we require that the number of individuals in each partition is greater than or equal to $2$, which implies that $m \geq 2$. Additionally, we require that $n \geq 2^s$ to ensure $m \geq 2$ in each stage.
    \item For the \((n, d, s)\) problem, the number of subgroups at each stage must be at least $2$. Otherwise, if we were to test the entire group as a single subgroup, it would make the process meaningless, since the detection result would already be known by the previous stage. To avoid certain unusual scenarios that arise when \( n \) is small, we focus on cases where \( n \geq d \cdot 2^s \).
    % \item The term m-partition refers to $P^m = \{ t_1 \geq t_2 \geq t_3 \geq \cdots \geq t_m \}$, where $t_i$ refers to the pool size respectively. Sometimes we will use $t_i$ to refer to the pool with $t_i$ individuals for convenience. In particular, $m$ average partition refers to the unique partition satisfying $t_1 = \lceil \frac{n}{m} \rceil$ and $t_m = \lfloor \frac{n}{m} \rfloor$, where $m$ and $n$ are given. It is easy to see that $P^m$ is an average partition if and only if $ 0 \leq t_1-t_m \leq 1$.
    \item The term \( m \)-partition refers to \( P^m = \{ t_1 \geq t_2 \geq t_3 \geq \cdots \geq t_m \} \), where \( t_i \) denotes the size of the \(i\)-th pool. For convenience, we may sometimes use \( t_i \) to directly refer to the pool containing \( t_i \) individuals. Specifically, an \( m \)-average partition is the unique partition such that \( t_1 = \lceil \frac{n}{m} \rceil \) and \( t_m = \lfloor \frac{n}{m} \rfloor \), given the values of \( m \) and \( n \). It is evident that \( P^m \) is an average partition if and only if \( 0 \leq t_1 - t_m \leq 1 \).
  \end{enumerate}

  % \begin{remark}
  %   Regarding point $3$ above, the requirement that \( n \geq d \cdot 2^s \) can be found in Algorithm $GTA_d$.
  % \end{remark}

  Next we will show a lemma about the monotonicity of $T_d(n,s)$ for $(n,d,s)$ problem.
  \begin{lemma} \label{mono}
      $T_d(n,s)$ is monotonically non-decreasing about $n$ and $d$.
  \end{lemma}
  \begin{proof}
    % It is known that every optimal algorithm for solving the $(n,d,s)$ problem is also an algorithm for solving the $(n^{\prime},d,s)$ problem, where $n\geq n^{\prime}$, although it may not be optimal for the latter problem. This is because the algorithm for $(n,d,s)$ can be modified to an algorithm for $(n^{\prime},d,s)$ by removing some unnecessary tests. Therefore, we have $T_d(n,s) \geq T_d(n^{\prime},s)$.

    % Similarly, any optimal algorithm for solving the $(n, d, s)$ problem can also be used to solve the $(n, d', s)$ problem, where $d \geq d^{\prime}$. Again, this is because the algorithm for $(n, d, s)$ can be modified to solve the $(n, d', s)$ problem by removing any redundant tests, resulting in $T_d(n,s) \geq T_{d'}(n,s)$.

  It is known that any optimal algorithm for solving the $(n, d, s)$ problem can also solve the $(n', d, s)$ problem for $n \geq n'$, though it may not remain optimal. This is because the original algorithm can be adapted by removing some unnecessary tests, implying that $T_d(n, s) \geq T_d(n', s)$.  

  Similarly, an optimal algorithm for the $(n, d, s)$ problem can be used to solve the $(n, d', s)$ problem when $d \geq d'$. Again, this adaptation involves eliminating redundant tests, leading to the inequality $T_d(n, s) \geq T_{d'}(n, s)$.  

  \end{proof}

  \begin{lemma} \label{taylor}
    Let $g(x) = \bigl( 1+x \bigr)^{\alpha} $ where $0 < \alpha < 1$ and $0 \leq x \leq 1$, then $g(x) \geq 1 + \alpha x - \frac{1}{8} x^2$.
  \end{lemma}
  \begin{proof}
    By Taylor expansion we know
    \begin{equation*}
      g(x)  \geq 1 + \alpha x + \frac{1}{2} \alpha(\alpha -1)x^2 \geq 1 + \alpha x - \frac{1}{8} x^2 > 0.
    \end{equation*}
  \end{proof}
  % Before presenting the testing algorithm for the $(n,d,s)$ problem, we first give the idea of Li in \cite{li1962sequential} that inspires us a lot in the following sections. At stage $1$, the $n$ items are arbitrarily divided into $g_1$ groups of $k_1$ (some possibly $k_1-1$) items. In general, at stage $i$, $2 \leq i \leq s$, items in defective subgroups are divided into $g_2$ groups of $k_2$ (some possibly $k_2-1$) items, $k_s$ is set to be $1$; thus every item is identified at stage $s$. Let $N_s$ denote the number of tests required by Li's $s$-stage algorithm. Roughly speaking, we have
  %   \begin{equation*}
  %     N_s = \sum_{i=1}^s g_i \leq \frac{n}{k_1}+\frac{d k_1}{k_2}+\cdots+\frac{d k_{s-2}}{k_{s-1}}+d k_{s-1} ,
  %   \end{equation*}
  %   ignoring the integral constraints, then the upper bound is minimized by
  %   \begin{equation*}
  %     k_i=\left(\frac{n}{d}\right)^{\frac{s-i}{s}} \quad  1 \leq i \leq s-1 .
  %   \end{equation*}
  %   This gives
  %   \begin{equation*}
  %     N_s \leq s d\left(\frac{n}{d}\right)^{\frac{1}{s}},
  %   \end{equation*}
  Before presenting our testing algorithm for the \((n,d,s)\) problem, we first discuss the foundational idea of Li \cite{li1962sequential}, which has provided valuable insights for our approach. At stage \(1\), the \(n\) items are arbitrarily divided into \(g_1\) groups, each containing \(k_1\) items (with some groups possibly having \(k_1 - 1\) items). In general, at stage \(i\) (\(2 \leq i \leq s\)), the items in the defective subgroups from the previous stage are further divided into \(g_i\) groups, each containing \(k_i\) items (with some groups possibly having \(k_i - 1\) items). At the final stage \(s\), \(k_s\) is set to \(1\), ensuring that each item is individually identified. We denote the total number of tests required by Li's \(s\)-stage algorithm as \(N_s\). Intuitively, we have
    \begin{equation*}
      N_s = \sum_{i=1}^s g_i \leq \frac{n}{k_1}+\frac{d k_1}{k_2}+\cdots+\frac{d k_{s-2}}{k_{s-1}}+d k_{s-1} ,
    \end{equation*}
  where we momentarily disregard the integer constraints for simplicity. The upper bound on \(N_s\) is minimized when
    \begin{equation*}
      k_i=\left(\frac{n}{d}\right)^{\frac{s-i}{s}} \quad  1 \leq i \leq s-1 .
    \end{equation*}
  This analysis yields the upper bound:
    \begin{equation*}
      N_s \leq s d\left(\frac{n}{d}\right)^{\frac{1}{s}},
    \end{equation*}
  indicating that \(O\left( sd \left(\frac{n}{d}\right)^{1/s}\right)\) tests are sufficient for the \((n,d,s)\) problem. The Algorithm \ref{algorithm1} and \ref{algorithmd} presented below represent an improvement and enhancement of the aforementioned strategy of Li\cite{li1962sequential}.

\section{Exact Solution of \texorpdfstring{$T_1(n,s)$}{Lg}  and \texorpdfstring{$T_d(n,2)$} {Lg} }
  % In this section, we provide the explicit expressions for \( T_1(n, s) \) and \( T_d(n, 2) \). For the $(n,1,s)$ problem, without loss of generality, let's suppose an $m$-partition occurs at the first stage. When the whole is divided into $m$ subgroups, only one subgroup is detected as positive. Given that the group testing problem is a minimax problem, our objective should be to minimize the size of the largest subgroup. Therefore, we can adopt an average $m$-partition strategy. This indicates that in the next $s-1$ stages, we consider the subgroup with $\lceil \frac{n}{m} \rceil$ individuals for testing.

  In this section, we provide the explicit expressions for \( T_1(n, s) \) and \( T_d(n, 2) \). For the \((n, 1, s)\) problem, without loss of generality, we assume an \( m \)-partition is performed at the first stage. When dividing the entire item into \( m \) subgroups, only one subgroup is detected as positive. As the group testing problem is framed as a minimax problem, our objective is to minimize the size of the largest subgroup. To this end, we employ an $m$ average partition strategy, ensuring each subgroup is as balanced in size as possible. Consequently, in the subsequent \( s - 1 \) stages, we focus on testing a subgroup of size \( \lceil \frac{n}{m} \rceil \).

  Consequently, we deduce the following recursive expression as Gajpal et al. \cite{gajpal2022optimal}:
  \begin{equation} \label{d=1-1}
    T_1^m(n,s) = m + T_1( \lceil \frac{n}{m} \rceil,s-1).
  \end{equation}
  To find the optimal solution for the $(n,1,s)$ problem, we need to minimize $T_1^m(n,s)$ over all possible values of $m$, which leads to the following expression:
  \begin{equation} \label{d=1-2}
    T_1(n,s) = \min_m T_1^m(n,s),
  \end{equation}
  and the boundary condition for this problem is given by:
  \begin{equation*}
    T_1(n,1) = n \quad (n \geq 2),
  \end{equation*}
  \begin{equation*}
    T_1(1,1) = \infty.
  \end{equation*}

  % In fact, it's hard to straight solve the nonlinear dynamic programming, however convenient to solve the dual problem of $(n,1,s)$ problem thus finding the largest $n$ such that the $(n,1,s)$ problem can be solved in fixed $t$ tests. Since it's divided equally at each stage, we can easily summarize the dual problem as an integer programming:
  In fact, directly solving the nonlinear dynamic programming problem proves to be quite challenging. However, it is convenient to tackle the dual problem of the \((n, 1, s)\) problem, which involves finding the largest \( n \) such that the \((n, 1, s)\) problem can be resolved using a fixed number of \( t \) tests. Since it's divided equally at each stage, we can readily formulate the dual problem as an integer programming problem:
  \begin{align*}
    &\max \quad \Pi_{i=1}^{s} m_i, \\
    &\ s.t. \quad \sum_{i=1}^{s} m_i = t.
  \end{align*}
  \begin{lemma} \label{mean}
    For positive integers $m_i(1\leq i \leq s)$ such $\sum_{i=1}^s m_i = st + i $, we claim that $\max \prod_{i=1}^s m_i = t^{s-i}(t+1)^{i}$.
  \end{lemma}
  \begin{proof}
    % To prove this, we first note that when $m_i-m_j \geq 2$, we can adjust $m_i^{\prime} = m_i - 1$, $m_j^{\prime} = m_j + 1$, and the rest $m_k^{\prime} = m_k$, which increases $\prod_{i=1}^s m_i$ to $\prod_{i=1}^s m_i^{\prime}$. Therefore, after a finite number of adjustments, when $\prod_{i=1}^s m_i$ is maximized, we know that the absolute difference between any two $m_i$ $(1\leq i \leq s)$ is at most one.

    % Next, we observe that since $\sum_{i=1}^s m_i = (s-i)t+i(t+1)$, $m_i$ can only take on values of $t$ or $t+1$. Hence, the maximum value of $\prod_{i=1}^s m_i$ is achieved when we take $s-i$ elements to be $t$ and $i$ elements to be $t+1$. Therefore, we have $\max \prod_{i=1}^s m_i = t^{s-i}(t+1)^i$, which completes the proof of the lemma.

    To prove this, we first note that when \( m_i - m_j \geq 2 \), we can adjust \( m_i^{\prime} = m_i - 1 \), \( m_j^{\prime} = m_j + 1 \), and leave the rest \( m_k^{\prime} = m_k \) unchanged. This adjustment increases \( \prod_{i=1}^s m_i \) to \( \prod_{i=1}^s m_i^{\prime} \). Therefore, after a finite number of such adjustments, the maximum product \( \prod_{i=1}^s m_i \) is achieved when the absolute difference between any two \( m_i \) (for \( 1 \leq i \leq s \)) is at most one.

    Next, we observe that since the sum \( \sum_{i=1}^s m_i = (s - i)t + i(t + 1) \), each \( m_i \) can only take values of \( t \) or \( t + 1 \). Hence, the maximum value of \( \prod_{i=1}^s m_i \) is achieved when we take \( s - i \) elements to be \( t \) and \( i \) elements to be \( t + 1 \). Therefore, we have \( \max \prod_{i=1}^s m_i = t^{s - i}(t + 1)^i \), which completes the proof of the lemma.
  \end{proof}

  % With Lemma \ref{mono} and Lemma \ref{mean}, we can immediately establish the following main theorem for $(n,1,s)$ problem:
  Using Lemma \ref{mono} and Lemma \ref{mean}, we can now derive the following main theorem for the $(n,1,s)$ problem:
  \begin{theorem} \label{main}
    For some integer $t$ and $i$ satisfying $ t^{s-i}(t+1)^{i} <  n \leq t^{s-i-1}(t+1)^{i+1}$ where $ 0 \leq i\leq s-1$, we have
    \begin{equation*} 
      T_1(n, s)=s t+i+1.
    \end{equation*}
  \end{theorem}

  % The proof of Lemma \ref{mean} actually provides us with an optimal algorithm, which we will refer to as Algorithm $GTA_1$. This algorithm works as follows:
  The proof of Lemma \ref{mean} not only establishes the theoretical result but also constructs an optimal algorithm, which we denote as Algorithm \ref{algorithm1}. The procedure operates as follows:
  \begin{algorithm}[!htbp]
    % \caption{$GTA_1$}
    \caption{}
      \label{algorithm1}
      \begin{algorithmic}
          \STATE step 1: Given an $(n,1,s)$ problem;
          \STATE step 2: Find the integer $t$ such that $t^s < n \leq (t+1)^s $;
          \STATE step 3: Find the integer $i$ where $  (0 \leq i \leq s-1) $ such that
          \begin{equation*}
            t^{s-i}(t+1)^{i} < n \leq t^{s-i-1}(t+1)^{i+1}.
          \end{equation*}
          \vspace{-1.2em}
          \STATE step 4: Divide the positive subgroup into $t+1$ equal parts at the first $i+1$ stages;
          \STATE step 5: Divide the positive subgroup into $t$ equal parts at the remaining $s-i-1$ stages.
      \end{algorithmic}
  \end{algorithm}

  We can also provide an alternative detection strategy as Algorithm \ref{algorithm2}.

    \begin{algorithm}[!htbp]
    % \caption{$GTA_2$}
    \caption{}
      \label{algorithm2}
      \begin{algorithmic}
          \STATE step 1: Given an $(n,d,s)$ problem;
          \STATE step 2: Find the integer $t$ such that $t^s < n \leq (t+1)^s $;
          % \STATE step 3: Find the integer $i$ where $  (0 \leq i \leq s-1) $ such that
          % \begin{equation*}
          %   t^{s-i}(t+1)^{i} < n \leq t^{s-i-1}(t+1)^{i+1}.
          % \end{equation*}
          % \vspace{-1.2em}
          \STATE step 3: Divide the positive subgroup into $t+1$ equal parts at the next stage and find out all positive subgroups;
          \STATE step 4: Repeat step $2$ and step $3$ for every positive subgroup with stage number $s-1$;
      \end{algorithmic}
  \end{algorithm}
  
  Actually the Algorithm \ref{algorithm2} is equivalent to Algorithm \ref{algorithm1} for $(n,1,s)$ problem, however the Algorithm \ref{algorithm2} is also applicable to general $(n,d,s)$ problem even though it may not be optimal.
  
  % By using Algorithm $GTA_1$, we can identify the unique defective item with the optimal number of tests. Next, we can obtain an important inequality estimation regarding $T_1(n,s)$.

  By employing Algorithm \ref{algorithm1}, we can reliably identify the unique defective item while achieving the theoretically optimal number of tests. Furthermore, we can obtain a key inequality bound for the function $T_1(n,s)$.

  \begin{proposition} \label{lower}
    $T_1(n,s) \geq s\sqrt[s]{n}$.
  \end{proposition}

  \begin{proof}
    Assuming $ t^{s-i}(t+1)^{i} < n \leq t^{s-i-1}(t+1)^{i+1}$ where $0 \leq i\leq s-1$, Theorem \ref{main} tells us that $T_1(n, s)=st+i+1$, we only need to check:
    \begin{equation*}
      st + i + 1 \geq s\sqrt[s]{t^{s-i-1}(t+1)^{i+1}},
    \end{equation*}
    which is equivalent to
    \begin{equation*} \label{ber}
      1+ \frac{i+1}{st} \geq (1+\frac{1}{t})^{\frac{i+1}{s}},
    \end{equation*}
    we can prove this using Bernoulli's inequality.
  \end{proof}

  % When the number of stages is only 1, the exact solution for the $(n,d,1)$ problem is trivially given by $T_d(n,1) = n$. In the case of $s=2$, we will continue our study to find the exact expression for $T_d(n,2)$, it's a cumbersome work but can provide some data and images, and the proof is left in the appendix.
  For the single-stage case ($s=1$), the $(n,d,1)$ problem admits a trivial exact solution with $T_d(n,1) = n$. When $s=2$, determining the exact expression for $T_d(n,2)$ becomes more involved. While this analysis is a cumbersome work, it yields valuable data and visualizations. For completeness, we provide the detailed proof in Appendix \ref{app1}.

  \begin{theorem} \label{s=2}
    For $(n,d,2)$ problem, and some integer $t \,(t\geq 2)$, $i$, $j$, and $n$ satisfying
    \begin{equation*}
      (dt+j)t^{1-i}(t+1)^i < n \leq (dt+j+1)t^{1-i}(t+1)^i,
    \end{equation*}
    where $ 0 \leq i\leq 1$ and $0 \leq j \leq d-1 $, we have
    \begin{equation*} 
      T_d(n,2) = 2dt + di + j + 1.
    \end{equation*}
  \end{theorem}

  \begin{remark}
    When $i = 0$ or $1$, it is easy to check that $t^{1-i}(t+1)^i=t+i$.
  \end{remark}

  \begin{proof}
    See Appendix \ref{app1} for the technical proof.
  \end{proof}
  % Next we will give a proportion of $T_d(n,2)$ which is similar to Proposition \ref{lower}
  We now present a bound for $T_d(n,2)$ that parallels Proposition \ref{lower} in structure and methodology.
  \begin{proposition} \label{lowers=2}
    $T_d(n,2) \geq 2d\sqrt{\frac{n}{d}}$.
  \end{proposition}
  \begin{proof}
    According to Theorem \ref{s=2}, when $(dt+j)(t+i) < n \leq (dt+j+1)(t+i)$, we have $T_d(n,2) = 2dt + di + j + 1$, hence
    \begin{equation*}
      2d\sqrt{\frac{n}{d}} \leq 2d\sqrt{\frac{(dt+j+1)(t+i)}{d}} = 2\sqrt{(dt+j+1)(dt+di)} \leq 2dt + di + j + 1.
    \end{equation*}
  \end{proof}

%%%%%%%%%%%%%%%%%%%%%%%%%%%%%%%%%%%%%%%%%%%%%%%%%%%%%%%%%%%%%%%%%%%%%%%%%%%%%%
\section{ Estimates of General \texorpdfstring{$T_d(n,s)$}{Lg} }

  \subsection{Dynamic Programming for General \texorpdfstring{$T_d(n,s)$}{Lg}}

  % We will now delve deeper into studying the optimal number of tests for general cases where $d$ ($d\geq 2$) and $s$ ($s\geq 3$) in the context of $(n,d,s)$ problem. After dividing the entire group into subgroups, there can be $1$, $2$, $\cdots$, or $d$ positive subgroups. However, we may not have knowledge about the number of positive individuals in each positive subgroup, which makes it challenging to directly apply dynamic programming.

  We now investigate the optimal number of tests for general cases of the $(n,d,s)$ problem where $d\geq 2$ and $s\geq 3$. When partitioning the population into subgroups, we encounter between $1$ and $d$ positive subgroups. However, the lack of knowledge regarding the exact number of positive individuals within each positive subgroup presents a significant challenge for direct application of dynamic programming approach.

  Considering a partition $P^m = \{ t_1 \geq t_2 \geq t_3 \geq \cdots \geq t_m \}$ with $t_1+t_2+t_3+\cdots+t_m=n$, after the initial group partition test, if there are $d$ defective individuals among $i$ defective subgroups, we consider the worst-case scenario, in this scenario, we assume that the $i$ largest subgroups, namely $t_1$, $t_2$, $\cdots$ , $t_i$ are tested positive, while the remaining subgroups are tested negative. To determine the optimal number of tests, we calculate the maximum number of tests for all possible values of $i$, and then take the minimum value across all possible partitions $P^m$. This generalized recurrence relation can be expressed as follows:

  \begin{equation}  \label{ditui}
    \begin{aligned}
      T_d^m(n,s) = m + &\min_{P^m} \Bigl( \max_{1\leq i\leq d} \bigl( G(t_1,t_2,\cdots,t_i,s-1) \big)\Bigr) , \\
    T_d(n,s) &= \min_{m} T_d^m(n,s).
    \end{aligned}
  \end{equation}

  Here $G(t_1,\cdots,t_i,s-1)$ represents the optimal number of tests when $t_l$ ($1\leq l \leq i$) subgroups are detected as positive and exactly contain $d$ defective individuals. Even though $G(t_1,\cdots,t_i,s-1)$ provides an optimal number, the specific testing number is not known unless certain special situations like $d=1$ or $d=2$. In particular, when $i=d$, it implies that each defective subgroup contains exactly one defective individual. In this case, we can observe that:
  \begin{equation*}
    G(t_1,\cdots,t_d,s-1) = T_1(t_1,s-1) + T_1(t_2,s-1) + \cdots + T_1(t_d,s-1).
  \end{equation*}

  For example, in the specific case of the $(n,3,4)$ problem, let's consider a partition $P^m = \{ t_1 \geq t_2 \geq t_3 \geq \cdots \geq t_m \}$. If $m \geq 3$, then when only subgroups $t_1$ and $t_2$ are tested positive, we cannot determine which subgroup contains exactly $2$ defective individuals. Consequently, we cannot obtain the numerical values from recursion, specifically the value of $G(t_1,t_2,3)$. As a result, the values of $T_3^m(n,4)$ and $T_3(n,4)$ are currently difficult to determine using recursion. This indicates that in the case of the $(n,3,4)$ problem and some similar situations, the recursive approach may not be sufficient to obtain the desired values. Alternative methods or strategies might be necessary to address these specific challenges and obtain the required information.

  \begin{remark}
    For the special case when $d=1$, we immediately observe  $G(t_1,s-1) = T_1(t_1,s-1) $. In this scenario,  the optimal partition $P^m$ is an \( m \)-average partition yielding  $t_1 = \lceil \frac{n}{m} \rceil$. Consequently, the recurrence relation \eqref{ditui} aligns precisely with the closed-form expressions given in \eqref{d=1-1} and \eqref{d=1-2}.

    For the case when \( d = 2 \), the recurrence relation \eqref{ditui} yields a concrete dynamic programming formulation \eqref{dy},
  \begin{equation} \label{dy}
    \begin{aligned}
      T_2^m(n,s) := m + \min_{ P^m} \Bigl\{ &\max \bigl \{ T_1(t_1,s-1) + T_1(t_2,s-1) , T_2(t_1,s-1) \bigr\} \Bigr\}, \\
    &T_2(n,s) = \min_{m} T_2^m(n,s),
    \end{aligned}
  \end{equation}
  this formulation enables the computation of exact values for \( T_2^m(n,s) \) and \(T_2(n,s) \).
  \end{remark}

  \subsection{Some Relevant Proposition}

  For a general $(n,d,s)$ problem, let's consider Algorithm \ref{algorithm2} to conduct and denote $H_d(n,s)$ as the testing number in the worst case. When $d=1$, it is straightforward to determine that $H_1(n,s) = T_1(n,s)$, when $d\geq 2$ we can get the following lemma.
  \begin{lemma} \label{pro1}
    $H_d(n,s) + d \leq dT_1(n,s)$ for $d \geq 2$.
  \end{lemma}
  \begin{proof}
    See Appendix \ref{app2} for the technical proof.
  \end{proof}

  When $m \geq d$, we consider the situation that partition $P^m$ is the average partition with $m$ subgroups. At this point, we specially define the quantity $U_d^m(n,s)$ as:

  \begin{equation*}
    U_d^m(n,s) := m + \max_{1\leq i\leq d} \left( G(t_1,\cdots,t_i,s-1) \right).
  \end{equation*}

  Here, $U_d^m(n,s)$ represents the minmax number of tests needed in the average partition with $m$ subgroups to identify $d$ defectives in the $(n,d,s)$ problem. Next we will show in $m\, (m \geq d)$ average partition case, the worst case is that $t_i\, (1 \leq i\leq d)$ subgroups exactly contain one defective. An important property of average partition testing is showed as follows. 
  \begin{lemma} \label{separation}
    \begin{equation*}
      U_d^m(n,s) = m + \sum_{i=1}^{d}\, T_1(t_i,s-1) \quad (m \geq d).
    \end{equation*}
  \end{lemma}
  \begin{proof}
    See Appendix \ref{app3} for the technical proof.
  \end{proof}

  \subsection{Estimate of Upper Bound}

  For the general $(n,d,s)$ problem, now we give a specific algorithm to find all defectives inspired by the strategy of Li \cite{li1962sequential}, correspondingly, we can provide an upper bound estimate of $T_d(n,s)$.
  \begin{theorem} \label{upbound}
    For some integer $t$, $i$, $j$ and $n$ satisfying
    \begin{equation*}
        (dt+j)t^{s-i-1}(t+1)^i < n \leq (dt+j+1)t^{s-i-1}(t+1)^i,
    \end{equation*}
    where $0 \leq i \leq s-1$ and $0 \leq  j \leq d-1 $, we can get
    \begin{equation*}
        T_d(n,s) \leq dst+di+j+1.
    \end{equation*}
  \end{theorem}
  \begin{proof}
    For $t$, $i$, $j$ and $n$ satisfying
    \begin{equation*}
      (dt+j)t^{s-i-1}(t+1)^i < n \leq (dt+j+1)t^{s-i-1}(t+1)^i,
    \end{equation*}
    at the first stage, we can just take $m=dt+j+1$ average partition, then according to Lemma \ref{separation}
    \begin{equation*}
        T^m_d(n,s) \leq U_d^m(n,s) = m + \sum_{i=1}^{d}\, T_1(t_i,s-1) \leq m + d T_1( \lceil \frac{n}{m} \rceil,s-1),
    \end{equation*}
    it's easy to check $\frac{n}{m} \leq t^{s-i-1}(t+1)^i $, and according to Theorem \ref{main} we obtain
    \begin{equation*}
      T_1( \lceil \frac{n}{m} \rceil,s-1) \leq (s-1)t+i,
    \end{equation*}
    therefore
    \begin{equation*}
      T_d(n,s) \leq T^m_d(n,s) \leq m + d \bigl( (s-1)t+i \bigr) = dst + di + j + 1.
    \end{equation*}
  \end{proof}

  From the proof of Theorem \ref{upbound}, for a general $(n,d,s)$ problem, we can devise the following detection strategy, denoted as Algorithm \ref{algorithmd}.

  \begin{algorithm}[!ht]
    % \caption{ $GTA_d$ }
    \caption{}
      \label{algorithmd}
      \begin{algorithmic}
          \STATE step 1: Given an $(n,d,s)$ problem;
          \STATE step 2: Find out the integer $t$ such that $dt^s < n \leq d(t+1)^s $;
          \STATE step 3: Find out the integer $i$ where $0 \leq i \leq s-1$ such that
          \begin{equation*}
            dt^{s-i}(t+1)^{i} < n \leq dt^{s-i-1}(t+1)^{i+1}.
          \end{equation*}
          \vspace{-1.2em}
          \STATE step 4:  Further find out the integer $j$ where $0 \leq  j \leq d-1 $ such that
          \begin{equation*}
            (dt+j)t^{s-i-1}(t+1)^i < n \leq (dt+j+1)t^{s-i-1}(t+1)^i.
          \end{equation*}
          \vspace{-1.2em}
          \STATE step 5: At the first testing stage, divide the whole into $m=dt+j+1$ average partitions;
          \STATE step 6: For all subgroups that are identified as positive at the first stage, use detection Algorithm \ref{algorithm2} at the remaining $s-1$ stages.
      \end{algorithmic}
  \end{algorithm}
  
  We denote $M_d(n,s)$ as the testing number for $(n,d,s)$ problem while taking Algorithm \ref{algorithmd}. Specially, according to Theorem $\ref{main}$ and Theorem $\ref{s=2}$, the Algorithm \ref{algorithmd} is the optimal algorithm and $M_d(n,s) = T_d(n,s)$ for $(n,1,s)$ and $(n,d,2)$ problem. For some integer $t$, $i$, $j$ and $n$ satisfying
  \begin{equation*}
      (dt+j)t^{s-i-1}(t+1)^i < n \leq (dt+j+1)t^{s-i-1}(t+1)^i,
  \end{equation*}
  we have $M_d(n,s) = dst+di+j+1 $ and we can obtain the following corollary:
  \begin{corollary} \label{dst}
    $M_d(dt^s,s)=sdt$.
  \end{corollary}
  For the general $(n,d,s)$ problem, we have the next theorem
  \begin{theorem} \label{uppower}
    For $ n \geq dt^s $ where $t \geq \lceil \frac{ds}{2} \rceil $, we can know
    \begin{equation*}
        M_d(n,s) < ds\bigl( \frac{n}{d} \bigr)^\frac{1}{s} + 2.
  \end{equation*}

  \end{theorem}
  \begin{proof}
    See Appendix \ref{app4} for the technical proof.
  \end{proof}

  \begin{corollary} \label{up2}
    For general $(n,d,s)$ problem, when $n \geq dt^s$ where $t \geq \lceil \frac{ds}{2} \rceil $, we have
    \begin{equation*}
      T_d(n,s) \leq M_d(n,s) \leq \lceil ds\bigl( \frac{n}{d} \bigr)^\frac{1}{s} \rceil + 1.
    \end{equation*}
  \end{corollary}

  \subsection{Estimate of Lower Bound}

  For a given $d$ ($d\geq 2$) and $s$ ($s\geq 3$) as well as a sufficiently large $n$, we now consider the lower bound of $T_d(n,s)$.
  \begin{lemma} \label{option}
    When $ n \geq n_1 := d\Big(\frac{e}{2} d^{2s-3}\Big)^s$, the first partition number $m$ where $m \leq d^{s-1}$ is not the best option for $(n,d,s)$ problem.
  \end{lemma}
  \begin{proof}
    See Appendix \ref{app5} for the technical proof.
  \end{proof}

  \begin{lemma} \label{derivation}
      When $m \geq d^{s-1}$, we define a new function for $t\in [\frac{n}{m},n]$
      \begin{equation*}
        f(t) := \sqrt[s-1]{t} + (d-1)\sqrt[s-1]{\frac{n-t}{m-1}},
      \end{equation*}
      then the minimum of $f(t)$ is $d \bigl( \frac{n}{m} \bigr) ^{\frac{1}{s-1}} $.
  \end{lemma}
  \begin{proof}
    See Appendix \ref{app6} for the technical proof.
  \end{proof}

  We next give an exact lower bound of $T_d(n,s)$ using last two lemmas.

  \begin{theorem} \label{avepartition}
    When $ n \geq d\Big(\frac{e}{2} d^{2s-3}\Big)^s $, for $(n,d,s)$ problem where $d \geq 2$ and $s \geq 3$, we have:
    \begin{equation*}
      T_d(n,s) \geq ds\bigl( \frac{n}{d} \bigr)^\frac{1}{s}.
    \end{equation*}
  \end{theorem}
  \begin{proof}
    See Appendix \ref{app7} for the technical proof.
  \end{proof}

  \begin{corollary} \label{lower2}
    According to Proposition \ref{lower} and Proposition \ref{lowers=2}, the inequality $T_d(n,s) \geq ds\bigl( \frac{n}{d} \bigr)^{1/s}$ holds on for $(n,1,s)$ and $(n,d,2)$ problem. Therefore when $ n \geq d\Big(\frac{e}{2} d^{2s-3}\Big)^s $, we can know for general $(n,d,s)$ problem $T_d(n,s) \geq ds\bigl( \frac{n}{d} \bigr)^{1/s}$, thus
    \begin{equation*}
      T_d(n,s) \geq \lceil ds\bigl( \frac{n}{d} \bigr)^\frac{1}{s} \rceil.
    \end{equation*}
  \end{corollary}

  \subsection{Optimal Estimation of upper and lower bounds}

  Above the text before, we have given the exact solution for $(n,1,s)$ and $(n,d,2)$ problem, and for the general $(n,d,s)$ problem, we can obtain the following final result.
  \begin{theorem} \label{last}
    For the general $(n,d,s)$ problem, when $n \geq d\Big(\frac{e}{2} d^{2s-3}\Big)^s$, we can get
      \begin{equation*}
        \lceil ds\bigl( \frac{n}{d} \bigr)^\frac{1}{s} \rceil \leq T_d(n,s) \leq \lceil ds\bigl( \frac{n}{d} \bigr)^\frac{1}{s} \rceil + 1,
      \end{equation*}
      here the upper bound and lower bound are both sharp.
  \end{theorem}

  \begin{proof}
    From a comprehensive perspective of Corollary \ref{up2} and Corollary \ref{lower2}, when $d=1$ or $s=2$, the inequality $\lceil ds\bigl( \frac{n}{d} \bigr)^\frac{1}{s} \rceil \leq T_d(n,s)$ is right, we only need to check $d\Big(\frac{e}{2} d^{2s-3}\Big)^s \geq d \bigl(\lceil \frac{ds}{2} \rceil \bigr)^s $ for $(n,d,s)$ problem when $d\geq 2$ and $s \geq 3$, it is easy to check
    \begin{equation*}
      \frac{e}{2} d^{2s-3}> \frac{d^{2s-3}}{2} > \frac{ds}{2} + 1,
    \end{equation*}
    hence the upper bound inequality holds on when $n \geq d\Big(\frac{e}{2} d^{2s-3}\Big)^s$.  
    
    To illustrate the optimality of the upper and lower bound, we claim $T_d(n,s)$ is possible to take the the upper and lower bound for sufficiently large $n$.
    For any $n =dt^s$ where $t$ is large enough, according to Corollary \ref{dst} we can know
    \begin{equation*}
      dst = \lceil ds\bigl( \frac{n}{d} \bigr)^\frac{1}{s} \rceil \leq T_d(n,s) \leq M_d(n,s)= M_d(dt^s,s) = dst,
    \end{equation*}
    hence $\lceil ds\bigl( \frac{dt^s}{d} \bigr)^\frac{1}{s} \rceil = T_d(dt^s,s)$ and the lower bound is sharp.
    
    As for the upper bound, specifically we only consider the $(n,1,3)$ problem, and the Algorithm \ref{algorithm1} give the exact value of $T_1(n,s)$, we claim there exists sufficiently large $n$ such that $T_1(n,3) - 1 > 3\sqrt[3]{n}$, let $n = t(t+1)^2+1$ where $t\geq 3$, and it is easy to check
    \begin{equation*}
      T_1(n,3) - 1=3t+2 > 3\sqrt[3]{t(t+1)^2+1},
    \end{equation*}
    hence $T_1(n,3) = \lceil 3n^{\frac{1}{3}} \rceil + 1$
    for large $n$. Above all, the bounds are both sharp.
  \end{proof}

\subsection{Numerical Experiments}

\begin{figure}[htbp]
      \centering
      \hspace{0em}
      \begin{minipage}[t]{0.49\textwidth}
      \centering
      \includegraphics[width=18em]{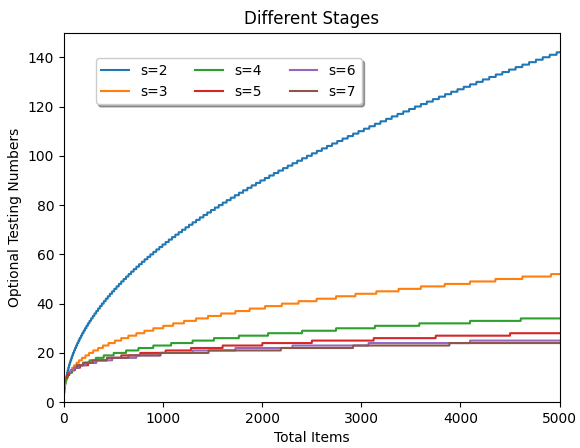}
      \caption{Defective $1$ for different stages}
      \label{d1}
      \end{minipage}
      \begin{minipage}[t]{0.49\textwidth}
      \centering
      \includegraphics[width=18em]{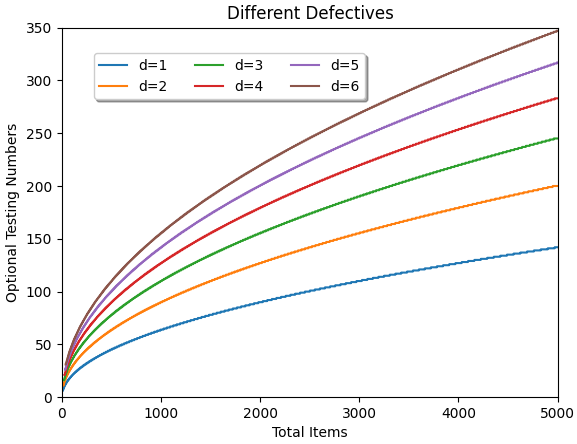}
      \caption{Stage $2$ for different defectives}
      \label{s2}
      \end{minipage}
  \end{figure}

  In this subsection, we will present numerical res ults and graphs of $T_d(n,s)$ for some fixed values of $d$ and $s$. Utilizing the exact expressions derived from Theorem $\ref{main}$ and Theorem $\ref{s=2}$, we have obtained the exact values of $T_1(n,s)$ for the $(n,1,s)$ problem and $T_d(n,2)$ for the $(n,d,2)$ problem. Furthermore, we will provide Figure \ref{d1}, which illustrates the values of $T_1(n,s)$ for $2 \leq s \leq 7$ and present Figure \ref{s2}, which displays the values of $T_d(n,2)$ for $1 \leq d \leq 6$.
  
\begin{figure}
    \centering
    \includegraphics[width=0.6\textwidth]{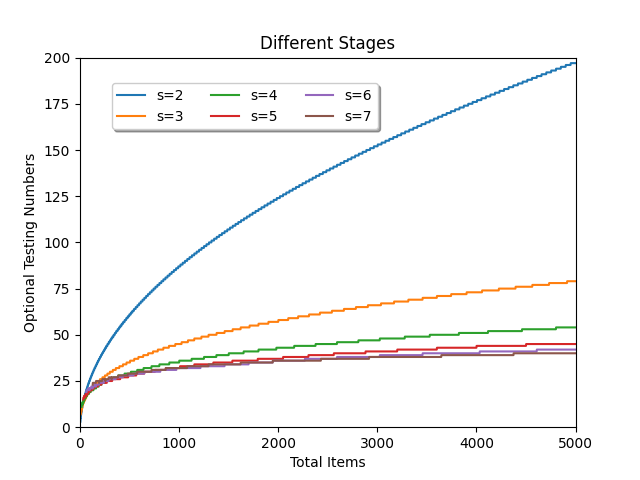}
    \caption{Defectives $2$ for different stages}
    \label{d2}
  \end{figure}

  Using recursion \eqref{dy}, we can calculate the specific values of $T_2(n,s)$  and plot Figure \ref{d2} for each $2 \leq s \leq 7$ and $n \leq 5000$. From these graphs, we can observe that the optimal number of tests aligns closely with the power function $ds\left(\frac{n}{d}\right)^{1/s}$, which is consistent with our previous estimation of upper and lower bounds. However, for the general $(n,d,s)$ problem, we cannot determine the value of $G(t_1,\cdots,t_i,s-1)$ through recursion, making it difficult to ascertain the value of $T_d^m(n,s)$. As a result, the specific values of $T_d(n,s)$ are uncertain, and we will not consider plotting figures for other $T_d(n,s)$ at this time. These numerical results and three graphs allow us to visualize the behavior of $T_d(n,s)$ for different settings of $d$ and $s$, providing valuable insights into the number of tests required to identify defectives

\section{Conclusions and Discussions}

  Our research lies on the MSGPT problem, where the subgroup tests within the same stage are disjoint and nonadaptive, while tests across different stages are adaptive. In the framework of $(n,d,s)$ problem, we obtain meaningful theoretical results. The solution relies fundamentally on two key components: the dynamic programming approach and the idea of average partition.

  For $(n,d,s)$ problems, previous results showed that
  \begin{equation*}
    \left(1-o(1)\right) ds \left(\frac{d-1}{d}\right)^{1-\frac{1}{s}} \left(\frac{n}{d}\right)^\frac{1}{s},
  \end{equation*}
  tests are necessary in Damaschke et al. \cite{damaschke2013two} and $\left(1+o(1)\right)ds\left(\frac{n}{d}\right)^{1/s}$ tests are sufficient in Li \cite{li1962sequential}. However, the upper bound and lower bound results are not sharp, we develop a new method via dynamic programming and obtain a much better result as shown in Theorem \ref{last}: when $n \geq d\Big(\frac{e}{2} d^{2s-3}\Big)^s$, we can achieve
  \begin{equation*}
    \lceil ds\bigl( \frac{n}{d} \bigr)^\frac{1}{s} \rceil \leq T_d(n,s) \leq \lceil ds\bigl( \frac{n}{d} \bigr)^{\frac{1}{s}} \rceil + 1,
  \end{equation*}
  where two bounds are sharp. As far as we know, it is the first result to give the sharp upper bound of general defectives in group testing frame.

  We have got $T_d(n,s) \geq ds\bigl( \frac{n}{d} \bigr)^{1/s}$ for sufficiently big $n$, but it's not a consistent lower bound for every $n$. Take $(513,2,4)$ problem for example, according to dynamic programming \eqref{dy}, we can get when $m=2$, $t_1=505$ and $t_2=8$
  \begin{equation*}
    T_2(n,4) = 2 + \max \bigl \{ T_1(505,3) + T_1(8,3) , T_2(505,3) \bigr\} = 32,
  \end{equation*}
  while $8\bigl( \frac{513}{2} \bigr)^{1/4} > 32$, which demonstrates that the lower bound $ds\left(\frac{n}{d}\right)^{1/s}$ is not consistent for all $(n,d,s)$ problems. In contrast to the general hypergeometric group testing problem $(d,n)$, which has a well known lower bound of $\lceil\log C_n^d\rceil$, the lower bound for the $(n,d,s)$ problem is not as straightforward. Therefore, it is important to consider both upper and lower bounds in order to obtain a more accurate understanding of the number of tests required in the MSGPT problem.

  Moreover, for sufficiently large $n$, if $M_d(n,s) = \lceil ds\left( \frac{n}{d} \right)^{1/s} \rceil$, we can consider Algorithm \ref{algorithmd} as the optimal algorithm for the $(n,d,s)$ problem. However, it is also possible that $M_d(n,s) = \lceil ds\left( \frac{n}{d} \right)^{1/s} \rceil + 1$, and in such cases, we cannot determine if Algorithm \ref{algorithmd} is optimal, except when $d=1$ or $s=2$. It's worth noting that the testing number $M_d(n,s)$ provided by Algorithm \ref{algorithmd} can take both $\lceil ds\left( \frac{n}{d} \right)^{1/s} \rceil$ and $\lceil ds\left( \frac{n}{d} \right)^{1/s} \rceil + 1$. Although we don't prove the optimality of Algorithm \ref{algorithmd} for all situations, personally, we believe that Algorithm \ref{algorithmd} is the optimal algorithm for sufficiently large values of $n$ such as $n \geq d\left(\frac{e}{2} d^{2s-3}\right)^s$. However, a general proof for all situations is currently lacking.

  The models and problems studied in this paper can be extended in several promising directions. First, our current analysis assumes the number of defectives \(d\) is known; however, if \(d\) is unknown but bounded by a fixed constant \(B\), it would be valuable to develop an efficient algorithm that performs well as \(n\) grows large. Second, while we have examined the MSGPT problem within the combinatorial paradigm, a natural extension would be to explore it in a probabilistic setting, building on existing works such as \cite{aldridge2019group,dorfman1943detection,finucan1964blood,samuels1978exact,sobel1966binomial}. Finally, generalizing our model to accommodate different noise patterns, an area of growing interest in information theory, as seen in \cite{scarlett2018noisy,scarlett2020noisy,teo2022noisy} could yield further insights.

%%%%%%%%%%%%%%%%%%%%%%%%%%%%%%%%%%%%%%%%%%%%%%%%%%%%%%%%%%%%%%%%%%%%%%%%%%%%%%%%%%%%%%%%%%
\appendix

\numberwithin{lemma}{section}

\section{Proof of Theorem \ref{s=2}} \label{app1}

\begin{proof}%[Proof of Theorem \ref{s=2}] 
  Consider an $m$-partition $P^m = \{ t_1 \geq t_2 \geq t_3 \geq \cdots \geq t_m \}$ with $t_1+t_2+t_3+\cdots+t_m=n$. If a subgroup is detected as positive at the first stage, then at the second stage all individuals need to be tested, so we should consider the worst situation and obtain an expression for $T_d^m(n,2)$ as follows:

  \begin{equation*}
    T_d^m(n,2) =
      \begin{cases}
        m + \min_{P^m} \sum_{i=1}^{d} t_i, \ & m \geq d ,\\
        m+n, & m < d.
      \end{cases}
  \end{equation*}

  Since the average partition can minimize $\sum_{i=1}^{d} t_i$, we only need to consider the cases that $P^m$ is an average partition. Let us assume $m \geq d$, and we claim that:
  \begin{equation*}
    \min_{m\geq d} T_d^m(n,2) = 2dt + di + j + 1,
  \end{equation*}
  to prove this, we need to show that ``$\leq$'' and ``$\geq$'' both hold.

  The ``$\leq$'' part is straightforward. We can take $m=dt+j+1$, since $n \leq (dt+j+1)t^{1-i}(t+1)^i$, we have $t_1 = \lceil \frac{n}{m} \rceil \leq t^{1-i}(t+1)^i = t+i$. It follows that:
  \begin{equation*}
    T_d(n,2) \leq T_d^m(n,2) = 2dt + di + j + 1.
  \end{equation*}

  For ``$\geq$'' part, let $m = dt+j+1+k \, (m \geq d)$ where $k$ maybe negative. We need to show that $T_d^m(n,2) \geq 2dt + di + j + 1 $, which means that we only need to prove that $\sum_{i=1}^{d} t_i \geq d(t+i) - k$. 
  
  We will use proof by contradiction, suppose that the inequality is not true, which implies that $\sum_{i=1}^{d} t_i \leq d(t+ i) - k-1$. We set $k = ad + b$ where $0 \leq b \leq d-1$ and $a$ may be negative. Then we have:
  \begin{equation*}
    \sum_{i=1}^{d} t_i \leq d(t+i) - k-1 = (t+i-a)d - b - 1.
  \end{equation*}
  From this inequality, we can see that $t_d \leq t+i-a-1$ and $t_i \leq t_d \leq t+i-a-1$ for those $d<i\leq m$. Hence
  \begin{align*}
    n &= \sum_{i=1}^{m} t_i = \sum_{i=1}^{d} t_i + \sum_{i=d+1}^{m} t_i \\
    &\leq d(t+i) - k-1 +(m-d)(t+i-a-1)  \\
    & = m(t+i) - (k+1) -(m-d)(a+1) \\
    & = (dt+j)(t+i) + (k+1)(t+i-1) -(m-d)(a+1),
    % & := (dt+j)(t+i) + h(t),
  \end{align*}
  for convenience, we define
  \begin{equation*}
      h(t) := (k+1)(t+i-1) -(m-d)(a+1),
  \end{equation*}
  we next illustrate $ h(t) \leq 0\, (t \geq 1)$  where $h(t)$ can be rewritten as
  \begin{equation*}
    h(t) = (k+1)(t+i-1) -(dt+j+1+k -d)(a+1) ,
  \end{equation*}
  consider $h(t)$ as a linear function of $t$, then the slope of the function is
  \begin{equation*}
    k+1-d(a+1)=ad+b+1-ad-d\leq 0,
  \end{equation*}
  so we can get
  \begin{equation*}
    h(t) \leq h(1) = (k+1)i - (j+1+k)(a+1) = (i-a-1)(k+1)-j(a+1),
  \end{equation*}
  if $a \geq 0$, easily $k \geq 0$  and $i-a-1 \leq 0$ so $h(1)\leq 0$, if $a=-1$, easily $k \leq -1$ so $h(1) = i(k+1) \leq 0$, elseif $a \leq -2$ then $k+d < 0$
  \begin{equation*}
    h(1) \leq - (j+1+k)(a+1) \leq j+1+k \leq d + k < 0,
  \end{equation*}
  hence $h(1) \leq 0$, therefore 
  \begin{equation*}
    n \leq  (dt+j)(t+i)  + h(t) \leq (dt+j)(t+i) = (dt+j)t^{1-i}(t+1)^i < n,
  \end{equation*}
  contradiction, so when $m \geq d$
  \begin{equation*}
    T_d^m(n,2) \geq 2dt + di + j + 1.
  \end{equation*}

  Now we consider the case when $m < d$. In this case, we have $ \min_{m\leq d} T_d^m(n,2) = 2+n$. It is easy to observe that when $t \geq 2$ and
  \begin{equation*}
    (dt+j)t^{1-i}(t+1)^i < n \leq (dt+j+1)t^{1-i}(t+1)^i,
  \end{equation*}
  for $m \geq d$, we can check
  \begin{equation*}
    \min_{m\geq d} T_d^m(n,2) = 2dt + di + j + 1 \leq (dt+j)(t+i) + 1 < 2+n.
  \end{equation*}
  Above all
  \begin{equation*}
    T_d(n,2) = \min_m T_d^m(n,2) = \min_{m\geq d} T_d^m(n,2) = 2dt + di + j + 1.
  \end{equation*}
\end{proof}

\section{Proof of Lemma \ref{pro1}} \label{app2}
\begin{proof}%[Proof of Lemma \ref{pro1}] 
  Let's proceed with a mathematical induction on the stage number $s$. The case $s=1$ is trivial, suppose the proposition $H_d(n,s-1) + d \leq dT_1(n,s-1)$ holds for some $d \geq 2$. This implies that $H_d(n,s-1) \leq dT_1(n,s-1)$ holds for $d \geq 1$. 

  Applying Algorithm \ref{algorithm2} to the $(n,d,s)$ problem, we can directly determine the number of defectives in each defective subgroup. This determination has no influence on the subsequent testing process. Following the initial partition $P^m = \{ t_1 \geq t_2 \geq t_3 \geq \cdots \geq t_m \}$ for the $(n,d,s)$ problem, suppose $t_i$ subgroups $(1\leq i \leq k)$ contain $d_i$ defectives, where $\sum_{i=1}^{k} d_i = d$ and $1\leq k \leq d$. According to the induction hypothesis, we have $H_{d_i}(t_i,s-1) \leq d_iT_1(t_i,s-1)$ for $1\leq i \leq k$. Now, we can extend this induction hypothesis to the $(n,d,s)$ problem as follows:
  \begin{align*}
      m + \sum_{i=1}^{k} H_{d_i}(t_i,s-1) + d & \leq m + d + \sum_{i=1}^{k} d_i T_{1}(t_i,s-1) \\
      & \leq m + d + \sum_{i=1}^{k} d_i T_{1}(t_1,s-1) \\
      & = m + d + dT_{1}(t_1,s-1) \\
      & \leq md + dT_{1}(t_1,s-1) \\
      & = dT_1(n,s) .
  \end{align*}
  Given any possible distribution of defectives $D = (d_1,d_2,d_3, \cdots, d_k)$, where $1\leq k \leq d$, the following inequality holds:
  \begin{equation*}
    m + \sum_{i=1}^{k} H_{d_i}(t_i,s-1) + d \leq dT_1(n,s).
  \end{equation*}
  By considering all possible distributions $D$, we can complete the induction by obtaining:
  \begin{equation*}
    H_d(n,s) =m + \max_{D} \sum_{i=1}^{k} H_{d_i}(t_i,s-1).
  \end{equation*}
  Therefore, we can conclude that $H_d(n,s) + d \leq dT_1(n,s)$ holds.
\end{proof}

\section{Proof of Lemma \ref{separation}} \label{app3}
\begin{proof}%[Proof of Lemma \ref{separation}] 
  After an $m$ average partition, there maybe appear $d$ or less than $d$ subgroups, we claim
  \begin{equation*}
    G(t_1,\cdots,t_i,s-1) \leq G(t_1,\cdots,t_d,s-1)  \quad  (i < d),
  \end{equation*}
  we take Algorithm \ref{algorithm2} to each positive subgroup at the following stages, this is the optimal strategy for $G(t_1,\cdots,t_d,s-1)$, while an upper bound for $G(t_1,\cdots,t_i,s-1)(i<d)$. 

  For $G(t_1,\cdots,t_i,s-1)(i<d)$, we can straight let $t_i$ subgroup contain $d_i$ defectives  with $\sum_{j=1}^i d_j = d$ for the reason that it has no influence in the determinant Algorithm \ref{algorithm2}. Without loss of generality, we can further set $d_j=1(1\leq j \leq k \leq i)$ and $d_j\geq 2( k < j \leq i)$ which implies $\sum_{j=k+1}^i d_j = d-k $, thus according to Lemma \ref{pro1}
  \begin{align*}
    G(t_1,\cdots,t_i,s-1) & = \sum_{j=1}^i H_{d_j}(t_j,s-1) \\
    & = \sum_{j=1}^k H_{d_j}(t_j,s-1) + \sum_{j=k+1}^i H_{d_j}(t_j,s-1)\\
    & \leq \sum_{j=1}^k T_{1}(t_j,s-1) + \sum_{j=k+1}^i \bigl(  d_j T_{1}(t_j,s-1) -d_j \bigr),
  \end{align*}
  notice that $t_1-t_d \leq 1$ for average partition, thus $ T_{1}(t_1,s-1)-T_{1}(t_d,s-1) \leq 1$, so
  \begin{align*}
    G(t_1,\cdots,t_i,s-1)& \leq \sum_{j=1}^k T_{1}(t_j,s-1) + \sum_{j=k+1}^i \bigl(  d_j T_{1}(t_d,s-1)  \bigr) \\
    &=  \sum_{j=1}^k T_{1}(t_j,s-1) + (d-k) \bigl(T_{1}(t_d,s-1)  \bigr) \\
    &\leq  \sum_{j=1}^k T_{1}(t_j,s-1) + \sum_{j=k+1}^d \bigl( T_{1}(t_j,s-1)  \bigr) \\
    & = \sum_{j=1}^d T_{1}(t_j,s-1),
  \end{align*}
  hence $ G(t_1,\cdots,t_i,s-1) \leq G(t_1,\cdots,t_d,s-1) $ holds on and this lemma is correct.
\end{proof}

\section{Proof of Theorem \ref{uppower}} \label{app4}
\begin{proof}%[Proof of Theorem \ref{uppower}] 
  For $t$, $i$, $j$ and $n$ satisfying
  \begin{equation*}
    (dt+j)t^{s-i-1}(t+1)^i < n \leq (dt+j+1)t^{s-i-1}(t+1)^i,
  \end{equation*}
  we have $M_d(n,s) = dst+di+j+1$. When $t \geq \frac{ds}{2}$, we claim
  \begin{equation*}
    ds\bigl( \frac{ (dt+j)t^{s-i-1}(t+1)^i }{d} \bigr)^\frac{1}{s} \geq dst + di + j - 1,
  \end{equation*}
  which is equivalent to
  \begin{equation*}
      (1+\frac{1}{t})^{\frac{i}{s}}(1+\frac{j}{dt})^{\frac{1}{s}} \geq 1+\frac{i}{st}+\frac{j-1}{dst},
  \end{equation*}
  by Lemma \ref{taylor}, we consider proving
  \begin{equation*}
      (1+\frac{i}{st}-\frac{1}{8t^2})(1+\frac{j}{dst}-\frac{1}{8t^2})\geq 1+\frac{i}{st}+\frac{j-1}{dst},
  \end{equation*}
  it's sufficient to prove
  \begin{equation*}
      \frac{1}{dst} \geq \frac{1}{8t^2} \bigl( 2 + \frac{i}{st} + \frac{j}{dst} \bigr),
  \end{equation*}
  thus sufficient to prove
  \begin{equation*}
    \frac{1}{dst} \geq \frac{1}{8t^2} \bigl( 2 + \frac{1}{t} + \frac{1}{st} \bigr),
  \end{equation*}
  when $t \geq \frac{ds}{2}$, easy to check
  \begin{equation*}
    \frac{1}{8t^2} \bigl( 2 + \frac{1}{t} + \frac{1}{st} \bigr) \leq \frac{1}{2t^2} \leq \frac{1}{dst},
  \end{equation*}
  hence $t \geq \frac{ds}{2}$ is a sufficient condition to get
  \begin{equation*}
    M_d(n,s) < ds\bigl( \frac{n}{d} \bigr)^\frac{1}{s}+ 2,
  \end{equation*}
  which is actually equivalent to
  \begin{equation*}
    M_d(n,s) \leq \lceil ds\bigl( \frac{n}{d} \bigr)^\frac{1}{s} \rceil + 1.
  \end{equation*}
\end{proof}

\section{Proof of Lemma \ref{option}} \label{app5}
\begin{proof}%[Proof of Lemma \ref{option}] 
  According to Theorem \ref{uppower}, when $n\geq n_1$, we know
  \begin{equation*}
    T_d(n,s) \leq ds\bigl( \frac{n}{d} \bigr)^\frac{1}{s}+ 2,
  \end{equation*}
  so we need to illustrate some lower bound of $T^m_d(n,s)(m\leq d^{s-1})$ is bigger than the upper bound $ds\bigl( \frac{n}{d} \bigr)^{1/s}+ 2$.

  We take an $m\,(m\geq 2)$ partition $P^m = \{ t_1 \geq t_2 \geq t_3 \geq \cdots \geq t_m  \}$ at first, after this testing and according to recurrence relation \eqref{ditui}, there exists one possible situation that $t_1$ and $t_2$ subgroups are conducted positive, according to Lemma \ref{mono}, it implies
  \begin{equation*}
    \max_{1\leq i\leq d} \bigl( G(t_1,\cdots,t_i,s-1) \big) \geq T_1(t_1,s-1) + T_1(t_2,s-1).
  \end{equation*}
  Combining the lower bound estimate of Proposition \ref{lower} for $T_1(n,s)$, we can easily see that
  \begin{align*}
    T^m_d(n,s) & \geq m+ \min_{P^m} \Bigl( T_1(t_1,s-1)+T_1(t_2,s-1)  \Bigr) \\ 
    & \geq m+(s-1) \min_{P^m} \bigl(\sqrt[s-1]{t_1} + \sqrt[s-1]{t_2}\bigr) \\
    & \geq m+ (s-1) \min_{P^m} \sqrt[s-1]{t_1+t_2}\\
    & \geq  m+ (s-1)  \sqrt[s-1]{\frac{ 2n }{m}} \\
    & \geq  m+ (s-1)  \sqrt[s-1]{\frac{ 2n }{d^{s-1}}},
  \end{align*}
  now we get a coarser lower bound of $T_d^m(n,s)$. When $n\geq n_2 := \Bigl(\frac{ds}{s-1}\Bigr)^{s(s-1)}\frac{d^{(s-1)^2}}{2^s}$, it's equivalent to
  \begin{equation*}
    (s-1)\sqrt[s-1]{\frac{2n}{d^{s-1}}} \geq ds\sqrt[s]{\frac{n}{d}},
  \end{equation*}
  hence when $n\geq n_2$, and according to Theorem \ref{uppower}, we have
  \begin{equation*}
    T^m_d(n,s) \geq m+ (s-1)  \sqrt[s-1]{\frac{ 2n }{d^{s-1}}} \geq 2+ds\sqrt[s]{\frac{n}{d}} > T_d(n,s).
  \end{equation*}
  Furthermore, we consider $n \geq n_1 > n_2$. In this case, we obtain the lemma that when $n \geq n_1$, the value of $m$ within the interval $[2, d^{s-1}]$ is not the optimal choice for the $(n, d, s)$ problem.
\end{proof}

\section{Proof of Lemma \ref{derivation}} \label{app6}
\begin{proof}%[Proof of Lemma \ref{derivation}] 
  Directly taking the derivation of $f(t)$
  \begin{equation*}
    f^\prime(t) = \dfrac{1}{s-1} \frac{d-1}{m-1} t^{\frac{1}{s-1}-1} [ \frac{m-1}{d-1} - (\frac{mt-t}{n-t})^{\frac{s-2}{s-1}}],
  \end{equation*}
  clearly $f(t)$ increases first and then decreases in the interval $[\frac{n}{m},n] $, in this case, the minimum of $f(t)$ within the range can only occur at the two endpoints, when $m \geq d^{s-1}$:
  \begin{equation*}
    f_{\min}(t) = \min \{ f(\frac{n}{m}),f(n) \} =\min \{ d\sqrt[s-1]{\frac{n}{m}}, \sqrt[s-1]{n} \} = d\sqrt[s-1]{\frac{n}{m}} .
  \end{equation*}
\end{proof}

\section{Proof of Theorem \ref{avepartition}} \label{app7}
\begin{proof}%[Proof of Theorem \ref{avepartition}] 
  Now we straight consider the case $m \geq d^{s-1}$ according to Lemma \ref{option}, when we take partition $P^m = \{ t_1 \geq t_2 \geq t_3 \geq \cdots \geq t_m  \}$, easily
  \begin{equation*}
    \max_{1\leq i\leq d} \bigl( G(t_1,\cdots,t_i,s-1) \big) \geq G(t_1,\cdots,t_d,s-1),
  \end{equation*}
  then we can get
  \begin{equation*}
    T_d^m(n,s) \geq m + \min_{P^m} G(t_1,\cdots,t_d,s-1),
  \end{equation*}
  where 
  \begin{equation*}
    G(t_1,\cdots,t_d,s-1) = T_1(t_1,s-1) + T_1(t_2,s-1) + \cdots + T_1(t_d,s-1),
  \end{equation*}
  according to Proposition \ref{lower}
  \begin{equation*}
    G(t_1,\cdots,t_d,s-1) \geq \sum_{i=1}^{d} (s-1) \sqrt[s-1]{t_i},
  \end{equation*}
  hence
  \begin{equation*}
    T_d^m(n,s) \geq m + \min_{P^m}\sum_{i=1}^{d} (s-1) \sqrt[s-1]{t_i},
  \end{equation*}
  we next consider which partitions can make $\sum_{i=1}^{d}\sqrt[s-1]{t_i}$ be smallest. The minimum when $t_i$ define on real positive numbers is not bigger than they define on positive positive integers, we straight consider real numbers case next.

  When $t_d$ is fixed, we can adjust $t_j (d <j\leq m)$ bigger to make $t_i (1 \leq i\leq d)$ smaller, so we can assume $t_d = t_{d+1} = \cdots = t_m$ in the partition $P^m$. Since the functions $\sqrt[s-1]{t}$ is a concave function, we can adjust $t_i(2 \leq i \leq d)$ smaller and $t_1$ bigger to make $\sqrt[s-1]{t_1} + \sqrt[s-1]{t_i}$ smaller unless $t_i = t_d(2 \leq i \leq d)$. Here $t_2 = t_3 = \cdots = t_d = \cdots = t_m$ is a necessary condition to make $\sum_{i=1}^{d}\sqrt[s-1]{t_i}$ be smallest, hence $t_1 + (m-1)t_d = n$ and we can get
  \begin{equation*}
    \sum_{i=1}^{d} \sqrt[s-1]{t_i} = \sqrt[s-1]{t_1} + (d-1) \sqrt[s-1]{t_d} = f(t_1),
  \end{equation*}
  where $t_1 \geq \frac{n}{m}$, according to Lemma \ref{derivation}
  \begin{equation*}
    \min_{P^m}\sum_{i=1}^{d} \sqrt[s-1]{t_i} \geq \min_{t\geq \frac{n}{m}}f(t) = f(\frac{n}{m}) = d\sqrt[s-1]{\frac{n}{m}},
  \end{equation*}
  hence
  \begin{equation*}
    T_d^m(n,s) \geq m + d(s-1)\sqrt[s-1]{\frac{n}{m}},
  \end{equation*}
  now we can achieve
  \begin{align*}
    T_d(n,s) & = \min_m T_d^m(n,s) \\
    & \geq \min_m \{ m + \min_{P^m} G(t_1,\cdots,t_d,s-1) \} \\
    & \geq \min_m\{ m + d(s-1)\sqrt[s-1]{\dfrac{n}{m}} \}\\
    & = \min_m\{ m + d \sqrt[s-1]{\dfrac{n}{m}} + \cdots + d \sqrt[s-1]{\dfrac{n}{m}} \}\\
    & \geq ds\sqrt[s]{\dfrac{n}{d}} .
  \end{align*}
\end{proof}

\section*{Acknowledgments}

The author is deeply grateful to the two anonymous reviewers for their careful reading of this paper and their invaluable contributions to this research. Special thanks go to Professor Zhiyi Tan, whose support is indispensable, from the initial selection of the topic to the final submission. The author would also like to express sincere gratitude to Professor Qi Zhang for his insightful advice on the writing, submission and revision process, as well as to Xingming Wang and Lishi Yu for reviewing the technical proofs and providing constructive feedback.

\bibliographystyle{amsplain}

\bibliography{revision}

\providecommand{\bysame}{\leavevmode\hbox to3em{\hrulefill}\thinspace}
\providecommand{\MR}{\relax\ifhmode\unskip\space\fi MR }
\providecommand{\MRhref}[2]{%
  \href{http://www.ams.org/mathscinet-getitem?mr=#1}{#2}
}
\providecommand{\href}[2]{#2}
\begin{thebibliography}{10}

\bibitem{aldridge2019group}
Matthew Aldridge, Oliver Johnson, Jonathan Scarlett, et~al., \emph{Group testing: an information theory perspective}, Foundations and Trends{\textregistered} in Communications and Information Theory \textbf{15} (2019), no.~3-4, 196--392.

\bibitem{allemann2013efficient}
Andreas Allemann, \emph{An efficient algorithm for combinatorial group testing}, Information Theory, Combinatorics, and Search Theory, Springer, 2013, pp.~569--596.

\bibitem{chan2011non}
Chun~Lam Chan, Pak~Hou Che, Sidharth Jaggi, and Venkatesh Saligrama, \emph{Non-adaptive probabilistic group testing with noisy measurements: Near-optimal bounds with efficient algorithms}, 2011 49th Annual Allerton Conference on Communication, Control, and Computing (Allerton), IEEE, 2011, pp.~1832--1839.

\bibitem{damaschke2013two}
Peter Damaschke, Azam~Sheikh Muhammad, and Eberhard Triesch, \emph{Two new perspectives on multi-stage group testing}, Algorithmica \textbf{67} (2013), no.~3, 324--354.

\bibitem{damaschke2014strict}
Peter Damaschke, Azam~Sheikh Muhammad, and G{\'a}bor Wiener, \emph{Strict group testing and the set basis problem}, Journal of Combinatorial Theory, Series A \textbf{126} (2014), 70--91.

\bibitem{de2005optimal}
Annalisa De~Bonis, Leszek Gasieniec, and Ugo Vaccaro, \emph{Optimal two-stage algorithms for group testing problems}, SIAM Journal on Computing \textbf{34} (2005), no.~5, 1253--1270.

\bibitem{dorfman1943detection}
Robert Dorfman, \emph{The detection of defective members of large populations}, The Annals of mathematical statistics \textbf{14} (1943), no.~4, 436--440.

\bibitem{du2000combinatorial}
Dingzhu Du, Frank~K Hwang, and Frank Hwang, \emph{Combinatorial group testing and its applications}, vol.~12, World Scientific.

\bibitem{d2014lectures}
Arkadii~G D'yachkov, \emph{Lectures on designing screening experiments}, arXiv preprint arXiv:1401.7505 (2014).

\bibitem{d1982bounds}
Arkadii~Georgievich D'yachkov and Vladimir~Vasil'evich Rykov, \emph{Bounds on the length of disjunctive codes}, Problemy Peredachi Informatsii \textbf{18} (1982), no.~3, 7--13.

\bibitem{d2016hypergraph}
Arkady~G D'yachkov, Ilya~V Vorobyev, NA~Polyanskii, and V~Yu Shchukin, \emph{On a hypergraph approach to multistage group testing problems}, 2016 IEEE International Symposium on Information Theory (ISIT), IEEE, 2016, pp.~1183--1191.

\bibitem{finucan1964blood}
HM~Finucan, \emph{The blood testing problem}, Journal of the Royal Statistical Society: Series C (Applied Statistics) \textbf{13} (1964), no.~1, 43--50.

\bibitem{gajpal2022optimal}
Yuvraj Gajpal, SS~Appadoo, Victor Shi, and Guoping Hu, \emph{Optimal multi-stage group partition for efficient coronavirus screening}, Annals of Operations Research (2022), 1--17.

\bibitem{hwang1972method}
Frank~K Hwang, \emph{A method for detecting all defective members in a population by group testing}, Journal of the American Statistical Association \textbf{67} (1972), no.~339, 605--608.

\bibitem{hwang2006pooling}
Frank Kwang-ming Hwang and Ding-zhu Du, \emph{Pooling designs and nonadaptive group testing: important tools for dna sequencing}, vol.~18, World Scientific, 2006.

\bibitem{katona1973combinatorial}
Gyula~OH Katona, \emph{Combinatorial search problems}, A survey of combinatorial theory, Elsevier, 1973, pp.~285--308.

\bibitem{kautz1964nonrandom}
William Kautz and Roy Singleton, \emph{Nonrandom binary superimposed codes}, IEEE Transactions on Information Theory \textbf{10} (1964), no.~4, 363--377.

\bibitem{li1962sequential}
Chou~Hsiung Li, \emph{A sequential method for screening experimental variables}, Journal of the American Statistical Association \textbf{57} (1962), no.~298, 455--477.

\bibitem{malyutov2013search}
Mikhail Malyutov, \emph{Search for sparse active inputs: A review}, Information Theory, Combinatorics, and Search Theory: In Memory of Rudolf Ahlswede (2013), 609--647.

\bibitem{samuels1978exact}
SM~Samuels, \emph{The exact solution to the two-stage group-testing problem}, Technometrics \textbf{20} (1978), no.~4, 497--500.

\bibitem{scarlett2018noisy}
Jonathan Scarlett, \emph{Noisy adaptive group testing: Bounds and algorithms}, IEEE Transactions on Information Theory \textbf{65} (2018), no.~6, 3646--3661.

\bibitem{scarlett2020noisy}
Jonathan Scarlett and Oliver Johnson, \emph{Noisy non-adaptive group testing: A (near-) definite defectives approach}, IEEE Transactions on Information Theory \textbf{66} (2020), no.~6, 3775--3797.

\bibitem{sobel1966binomial}
Milton Sobel and Phyllis~A Groll, \emph{Binomial group-testing with an unknown proportion of defectives}, Technometrics \textbf{8} (1966), no.~4, 631--656.

\bibitem{teo2022noisy}
Bernard Teo and Jonathan Scarlett, \emph{Noisy adaptive group testing via noisy binary search}, IEEE Transactions on Information Theory \textbf{68} (2022), no.~5, 3340--3353.

\bibitem{wolf1985born}
Jack Wolf, \emph{Born again group testing: Multiaccess communications}, IEEE Transactions on Information Theory \textbf{31} (1985), no.~2, 185--191.

\end{thebibliography}

\end{document}